\documentstyle [12pt,epsf]{article}

\input{epsf}
\begin{document}
\begin{titlepage}
\begin{center}

 \vspace{-0.7in}

{\large \bf Stochastic Quantization of Scalar Fields\\in Einstein and Rindler Spacetime}\\
 \vspace{.3in}{\large\em G. Menezes\,\footnotemark[1]
and N.~F.~Svaiter
\footnotemark[2]}\\
\vspace{.3in}
Centro Brasileiro de Pesquisas F\'{\i}sicas\,-CBPF,\\
 Rua Dr. Xavier Sigaud 150,\\
 Rio de Janeiro, RJ, 22290-180, Brazil \\

\subsection*{\\Abstract}
\end{center}
\baselineskip .18in

We consider the stochastic quantization method for scalar fields
defined in a curved manifold and also in a flat space-time with
event horizon. The two-point function associated to a massive
self-interacting scalar field is evaluated, up to the first order
level in the coupling constant $\lambda$, for the case of an
Einstein and also a Rindler Euclidean metric, respectively. Its
value for the asymptotic limit of the Markov parameter
$\tau\rightarrow\infty$ is exhibited. The divergences therein are
taken care of by employing a covariant stochastic
regularization, where all the symmetries of the original theory are preserved.\\
\vspace{0,3cm}\\
PACS numbers: 03.70+k, 04.62.+v

\footnotetext[1]{e-mail:\,\,gsm@cbpf.br}
\footnotetext[2]{e-mail:\,\,nfuxsvai@cbpf.br}

\end{titlepage}
\newpage\baselineskip .20
in
\section{Introduction}
$\,\,\,\,\,\,\,$

\quad In the last century Parisi and Wu introduced the stochastic
quantization method \cite{parisi}. The main idea of the stochastic
quantization is that a $d$-dimensional quantum system is
equivalent to a $(d+1)$-dimensional classical system including
fluctuations effects into considerations. Some of the most
important papers in the subject can be found in Ref. \cite{ii}. A
brief introduction to the stochastic quantization can be found in
the Refs. \cite{namiki1} and \cite{sakita} and a complete review
of the stochastic quantization is given in the Ref. \cite{damre}.

This program of stochastic quantization and the stochastic
regularization was carried out for generic fields defined in flat,
Euclidean manifolds. It is important to observe that, since the
stochastic regularization is not an action regularization, this
method preserves all the symmetries of the theory under study. In
the development of this program some authors applied this method to
linearized Euclidean gravity and also non-linearized gravity
\cite{gravity1} \cite{grav1} \cite{grav2} \cite{grav3}
\cite{gravity2}. It is worth pointing out here that we cannot expect
that the method will improve the perburbative non-renormalizability
of quantum gravity. Working along such lines, we may observe that
the study of a  situation which lies between these two extremes is
missing. Therefore, before the attempt to implement the program in
non-renormalizable models, a consistent logical step is to discuss
an intermediate situation between fields in flat spacetime and
quantum gravity, i.e., the semiclassical theory, which corresponds
to the one-loop approximation to the totally quantized theory. It is
tempting to think that many unsolved problems, as the loss of
information in black hole formation and the entropy of black holes,
can gain new insights by introducing the stochastic quantization.

Although there are some papers discussing the stochastic
quantization in Minkowski spacetime \cite{con1} \cite{gozzi}, the
stochastic quantization was originally introduced as an
alternative quantization method for fields defined in Euclidean
space. Therefore our aim in this article is to discuss the
stochastic quantization of scalar fields defined in a static
curved manifold without event horizon and also in a flat manifold
with event horizon i.e., to analytically continue metrics that
generate real actions. We would like to stress that the stochastic
quantization is quite different from the other quantization
methods, therefore it can reveal new structural elements of a
theory which so far have gone unnoticed.

The method of stochastic quantization in flat spacetime with
trivial topology can be summarized by the following steps. First,
starting from a field defined in Minkowski spacetime, after
analytic continuation to imaginary time, the Euclidean
counterpart, i.e., the field defined in an Euclidean space, is
obtained. Second, it is introduced a monotonically crescent Markov
parameter, called in the literature "fictitious time" and also a
random noise field $\eta(\tau,x)$, which simulates the coupling
between the classical system and a heat reservoir. It is assumed
that the fields defined at the beginning in a $d$-dimensional
Euclidean space also depends on the Markov parameter, therefore
the field and a random noise field are defined in a
$(d+1)$-dimensional manifold. One starts with the system out of
equilibrium at an arbitrary initial state. It is then forced into
equilibrium assuming that its evolution is governed by a Markovian
Langevin equation with a white random noise field \cite{kac}
\cite{z} \cite{kubo}. In fact, this evolution is described by a
process which is stationary, Gaussian and Markovian. Finally, the
$n$-point correlation functions of the theory in the
$(d+1)$-dimensional space are defined by performing averages over
the random noise field with a Gaussian distribution, that is,
performing the stochastic averages
$\langle\,\varphi(\tau_{1},x_{1})
\varphi(\tau_{2},x_{2})...\varphi(\tau_{n},x_{n})\,\rangle_{\eta}$.
The $n$-point Schwinger functions of the Euclidean $d$-dimensional
theory are obtained  evaluating these $n$-point stochastic
averages $\langle\,\varphi(\tau_{1},x_{1})
\varphi(\tau_{2},x_{2})...\varphi(\tau_{n},x_{n})\,\rangle_{\eta}$
when the Markov parameter goes to infinity
$(\tau\rightarrow\infty)$, and the equilibrium is reached. This
can be proved in different ways for the particular case of
Euclidean scalar field theory. One can use, for instance, the
Fokker-Planck equation \cite{fp} \cite{ili} associated with the
equations describing the stochastic dynamic of the system. A
diagrammatical technique \cite{grimus} has also been used to prove
such equivalence.

The original method proposed by Parisi and Wu was extended to
include theories with fermions \cite{f1} \cite{f2} \cite{f3}. The
first question that appears in this context is if make sense the
Brownian problem with anticommutating numbers. It can be shown
that, for massless fermionic fields, there will not be a
convergence factor after integrating the Markovian Langevin
equation. Therefore the equilibrium is not reached. One way of
avoiding this problem is to introduce a kernel in the Langevin
equation describing the evolution of two Grassmannian fields.

Usually, the Parisi-Wu scheme of quantization applied to bosonic
and fermionic fields converges towards non-regularized theories.
In order to obtain regularized stochastic diagrams, the original
stochastic process proposed by Parisi and Wu must be modified. One
way is to replace the Markovian process by a non-Markovian one
\cite{fox1} \cite{fox2}. This can be done introducing a colored
noise in the Einstein relations \cite{bgz} \cite{zbern}
\cite{alfaro} \cite{iengo}. Recently, the stochastic quantization
of a topological Chern-Simons theory and the self-interacting
$\lambda\varphi^{4}$ theory was investigated by the authors, using
the Parisi-Wu scheme with a non-Markovian Langevin equation
\cite{menezes1} \cite{menezes2}.

Motivated by the problems that appear when one tries to implement
this regularization scheme in non-abelian gauge theories with
Zwanziger's gauge fixing \cite{zwanziger} \cite{bh} \cite{jiro},
Bern et al \cite{taubes} derived a truly non-perturbative
regularization procedure in $QCD$, still in the Markovian
scenario. They introduced a modified Langevin equation with a
regulator multiplying the noise. Therefore, the original Einstein
relations and consequently the noise probability distribution are
maintained. This continuum regularization program was very
successful in the invariant non-perturbative regularization of all
quantum field theories \cite{c1} \cite{c2} \cite{halp}, including
gauge theory, gauge theory with fermions and gravity. In few
words, the idea of any stochastic regularization is to start from
an interacting theory, then construct Langevin tree-graphs, where
each leg of which ends in a regularized noise factor. Since it is
possible to obtain the loops of the theory by contracting the
noise factors, one ends up with a theory where every closed loop
contains at least some power of the regulator. With this
modification, one can show that the system converges towards a
regularized theory. The next step to construct a finite theory is
to use for example a minimal-subtraction scheme in which the
ultraviolet divergent contributions are eliminated.

The aim of this paper is to implement the stochastic quantization
for the self-interacting $\lambda\varphi^{4}$ theory in a Riemannian
manifold and also in a flat manifold with event horizon, being more
specific, the Einstein \cite{eddington} and the Rindler spacetime
\cite{rin} \cite{full1} \cite{full2} \cite{sciama}, where
positive-definite forms can be obtained which generate real actions
in the Parisi-Wu Langevin equation. The only article that has some
overlap with our work is the Huang paper \cite{huang}. This author
have used the stochastic quantization to find the regularized
stress-energy tensor associated to a scalar field in an
inhomogeneous spacetime. We would like to stress that although
Brownian motion on a manifold has been studied by the
mathematicians, there are a few papers studying Brownian motion in
general manifolds. See for example the Ref. \cite{brow}.

This quantization method differs from the others, the canonical
and the path integral field quantization, based in the Hamiltonian
and the Lagrangian, respectively, in many aspects. The method
starts from a classical equation of motion, but not from
Hamiltonian or Lagrangian, and consequently can be used to
quantize dynamical systems without canonical formalism.
Furthermore, it is useful in situations where the others methods
lead to difficult problems and can bring us new important results.

For example, a quite important point in a regularization procedure
is that it must preserve all the symmetries of the unregularized
Lagrangian. Many authors have been stressed that a priori we can not
expect that a regularization independent proof of the
renormalization of theories in a curved background exists. The
presence of the Markov parameter as an extra dimension lead us to a
regularization scheme, which preserves all the symmetries of the
theory under study. Since the stochastic regularization is not an
action regularization, it may be a way to construct such proof. As a
starting point of this program, we should calculate the two-point
function up to the first order level in the coupling constant
$\lambda$ and apply the continuum stochastic regularization. Our
results are to be compared with the usual ones in the literature.

The organization of the paper is the following: in section II we
discuss the stochastic quantization for the
$(\lambda\varphi^{4})_{d}$ scalar theory in a $d$-dimensional
Euclidean manifold. In section III we use the stochastic
quantization and the stochastic regularization to obtain the
two-point Schwinger function in the one-loop approximation in the
Einstein manifold using the stochastic quantization. In section IV
we repeat the method for the case of the Rindler manifold, a
non-simple connected manifold in the Euclidean version.
Conclusions are given in the section V. In the appendix we sketch
the general formalism to construct the maximal analytic extension
of the vacuum Schwarzschild solution. In this paper we use
$\hbar=c=k_{B}=G=1$.

\section{Stochastic quantization for the
$(\lambda\varphi^{4})_{d}$ scalar theory: the Euclidean case}

In this section, we give a brief survey for the case of
self-interacting scalar fields, implementing the stochastic
quantization and the continuum stochastic regularization theory up
to the one-loop level. Let us consider a neutral scalar field with
a $(\lambda\varphi^{4})$ self-interaction. The Euclidean action
that usually describes a free scalar field is
\begin{equation}
S_{0}[\varphi]=\int d^{d}x\, \left(\frac{1}{2}
(\partial\varphi)^{2}+\frac{1}{2}
m_{0}^{2}\,\varphi^{2}(x)\right), \label{9}
\end{equation}
and the interacting part, defined by the non-Gaussian contribution,
is
\begin{equation}
S_{I}[\varphi]= \int d^{d}x\,\frac{\lambda}{4!} \,\varphi^{4}(x).
\label{10}
\end{equation}

The simplest starting point of the stochastic quantization to
obtain the Euclidean field theory is a Markovian Langevin
equation. Assume a flat Euclidean $d$-dimensional manifold, where
we are choosing periodic boundary conditions for a scalar field
and also a random noise. In other words, they are defined in a
$d$-torus $\Omega\equiv\,T\,^d$. To implement the stochastic
quantization we supplement the scalar field $\varphi(x)$ and the
random noise $\eta(x)$ with an extra coordinate $\tau$, the Markov
parameter, such that $\varphi(x)\rightarrow \varphi(\tau,x)$ and
$\eta(x)\rightarrow \eta(\tau,x)$. Therefore, the fields and the
random noise are defined in a domain: $T\,^{d}\times R\,^{(+)}$.
Let us consider that this dynamical system is out of equilibrium,
being described by the following equation of evolution:
\begin{equation}
\frac{\partial}{\partial\tau}\varphi(\tau,x)
=-\frac{\delta\,S_0}{\delta\,\varphi(x)}|_{\varphi(x)=\varphi(\tau,\,x)}+
\eta(\tau,x), \label{23}
\end{equation}
where $\tau$ is a Markov parameter, $\eta(\tau,x)$ is a random
noise field and $S_0$ is the usual free Euclidean action defined
in Eq.(\ref{9}). For a free scalar field, the Langevin equation
reads
\begin{equation}
\frac{\partial}{\partial\tau}\varphi(\tau,x)
=-(-\Delta+m^{2}_{0}\,)\varphi(\tau,x)+ \eta(\tau,x),
 \label{24}
\end{equation}
where $\Delta$ is the $d$-dimensional Laplace operator. The
Eq.(\ref{24}) describes a Ornstein-Uhlenbeck process and we are
assuming the Einstein relations, that is:
\begin{equation}
\langle\,\eta(\tau,x)\,\rangle_{\eta}=0, \label{28}
\end{equation}
and for the two-point correlation function associated with the
random noise field
\begin{equation}
\langle\, \eta(\tau,x)\,\eta(\tau',x')\,
\rangle_{\eta}\,=2\delta(\tau-\tau')\,\delta^{d}(x-x'), \label{29}
\end{equation}
where $\langle\,...\rangle_{\eta}$ means stochastic averages. The
above equation defines a delta-correlated random process. In a
generic way, the stochastic average for any functional of
$\varphi$ given by $F[\varphi\,]$ is defined by
\begin{equation}
\langle\,F[\varphi\,]\,\rangle_{\eta}=
\frac{\int\,[d\eta]F[\varphi\,]\exp\biggl[-\frac{1}{4} \int d^{d}x
\int d\tau\,\eta^{2}(\tau,x)\bigg]}
{\int\,[d\eta]\exp\biggl[-\frac{1}{4} \int d^{d}x \int
d\tau\,\eta^{2}(\tau,x)\bigg]}. \label{36}
\end{equation}
Let us define the retarded Green function for the diffusion problem
that we call $G(\tau-\tau',x-x')$. The retarded Green function
satisfies $G(\tau-\tau',x-x')=0$ if $\tau-\tau'<0$ and also
\begin{equation}
\Biggl[\frac{\partial}{\partial\tau}+(-\Delta_{x}+m^{2}_{0}\,)
\Bigg]G(\tau-\tau',x-x')=\delta^{d}(x-x')\delta(\tau-\tau').
\label{25}
\end{equation}
Using the retarded Green function and the initial condition
$\varphi(\tau,x)|_{\tau=0}=0$, the solution for Eq.(\ref{24}) reads
\begin{equation}
\varphi(\tau,x)=\int_{0}^{\tau}d\tau'\int_{\Omega}d^{d}x'\,
G(\tau-\tau',x-x')\eta(\tau',x'). \label{26}
\end{equation}
Let us define the Fourier transforms for the field  and the noise
given by $\varphi(\tau,k)$ and $\eta(\tau,k)$. We have respectively
\begin{equation}
\varphi(\tau,k)=\frac{1}{(2\pi)^\frac{d}{2}} \int\,d^{d}x\,
e^{-ikx}\,\varphi(\tau,x), \label{33}
\end{equation}
and
\begin{equation}
\eta(\tau,k)=\frac{1}{(2\pi)^\frac{d}{2}} \int\,d^{d}x\,
e^{-ikx}\,\eta(\tau,x). \label{34}
\end{equation}
Substituting Eq.(\ref{33}) in Eq.(\ref{9}), the free action for the
scalar field in the $(d+1)$-dimensional space writing in terms of
the Fourier coefficients reads
\begin{equation}
S_{0}[\varphi(k)]\,|_{\varphi(k)=\varphi(\tau,\,k)}=
\frac{1}{2}\int\,d^{d}k\,\varphi(\tau,k)(k^{2}+m_{0}^{2})\varphi(\tau,k).
\label{35}
\end{equation}
Substituting Eq.(\ref{33}) and Eq.(\ref{34}) in Eq.(\ref{24}) we
have that each Fourier coefficient satisfies a Langevin equation
given by
\begin{equation}
\frac{\partial}{\partial\tau}\varphi(\tau,k)
=-(k^{2}+m^{2}_{0})\varphi(\tau,k)+ \eta(\tau,k). \label{36}
\end{equation}
In the Langevin equation the particle is subject to a fluctuating
force (representing a stochastic environment), where its average
properties are presumed to be known and also the friction force.
Note that the "friction coefficient" in the Eq.(\ref{36}) is given
by $(k^{2}+m^{2}_{0})$.

The solution for Eq.(\ref{36}) reads
\begin{equation}
\varphi(\tau,k)=\exp\left(-(k^{2}+m_{0}^{2})\tau\right)\varphi(0,k)+
\int_{0}^{\tau}d\tau'\exp\left(-(k^{2}+m_{0}^{2})(\tau-\tau')\right)
\eta(\tau',k). \label{37}
\end{equation}
Using the Einstein relation, we get that the Fourier coefficients
for the random noise satisfies
\begin{equation}
\langle\,\eta(\tau,k)\,\rangle_{\eta}=0 \label{38}
\end{equation}
and
\begin{equation}
\langle\,\eta(\tau,k)\eta(\tau',k')\,\rangle
_{\eta}=2\delta(\tau-\tau')\delta^{d}(k+k'). \label{39}
\end{equation}
It is possible to show that
$\langle\,\varphi(\tau,k)\varphi(\tau',k')\,\rangle_{\eta}|_{\tau=\tau'}\equiv
D(k,k';\tau,\tau')$ is given by:
\begin{equation}
D(k;\tau,\tau)=(2\pi)^d\delta^{d}(k+k')\frac{1}{(k^{2}+m_{0}^{2})}\biggl(1-\exp\left(-2\tau
(k^{2}+m_{0}^{2})\right)\biggr). \label{44}
\end{equation}
where we assume $\tau=\tau'$.

Now let us analyze the stochastic quantization for the
$(\lambda\varphi^{4})_{d}$ self-interaction scalar theory. In this
case the Langevin equation reads
\begin{equation}
\frac{\partial}{\partial\tau}\varphi(\tau,x)
=-(-\Delta+m^{2}_{0}\,)\varphi(\tau,x)-\frac{\lambda}{3!}\varphi^{3}(\tau,x)+
\eta(\tau,x). \label{35}
\end{equation}
The two-point correlation function associated with the random field
is given by the Einstein relations, while the other connected
correlation functions vanish, i.e.,
\begin{equation}
\langle\,\eta(\tau_{1},x_{1})\eta(\tau_{2},x_{2})...\eta(\tau_{2k-1},
x_{2k-1})\,\rangle_{\eta}=0, \label{377}
\end{equation}
and also
\begin{equation}
\langle\eta(\tau_{1},x_{1})...\eta(\tau_{2k},x_{2k})\,\rangle_{\eta}=
\sum\,\langle\eta(\tau_{1},x_{1})\eta(\tau_{2},x_{2})\,\rangle_{\eta}
\langle\,\eta(\tau_{k},x_{k})\eta(\tau_{l},x_{l})\,\rangle_{\eta}...,
\label{388}
\end{equation}
where the sum is to be taken over all the different ways in which
the $2k$ labels can be divided into $k$ parts, i.e., into $k$
pairs. Performing Gaussian averages over the white random noise,
it is possible to prove the important formulae
\begin{equation}
\lim_{\tau\rightarrow\infty}
\langle\,\varphi(\tau_{1},x_{1})\varphi(\tau_{2},x_{2})...
\varphi(\tau_{n},x_{n}) \,\rangle_{\eta}= \frac{\int
[d\varphi]\varphi(x_{1})\varphi(x_{2})...\varphi(x_{n})
\,e^{-S(\varphi)}} {\int [d\varphi]\,e^{-S(\varphi)}}, \label{399}
\end{equation}
where $S(\varphi)=S_0(\varphi)+S_{I}(\varphi) $ is the
$d$-dimensional action. This result leads us to consider the
Euclidean path integral measure a stationary distribution of a
stochastic process. Note that the solution of the Langevin equation
needs a given initial condition. As for example
\begin{equation}
\varphi(\tau,x)|_{\tau=0}=\varphi_{0}(x). \label{40}
\end{equation}

Let us use the Langevin equation to perturbatively solve the
interacting field theory. One way to handle the Eq.(\ref{35}) is
with the method of Green's functions. We defined the retarded Green
function for the diffusion problem in the Eq.(\ref{25}). Let us
assume that the coupling constant is a small quantity. Therefore to
solve the Langevin equation in the case of a interacting theory we
use a perturbative series in $\lambda$. Therefore we can write
\begin{equation}
\varphi(\tau,x)=\varphi^{(0)}(\tau,x)+\lambda\varphi^{(1)}(\tau,x)+
\lambda^{2}\varphi^{(2)}(\tau,x)+... \label{41}
\end{equation}
Substituting the Eq.(\ref{41}) in the Eq.(\ref{35}), and if we
equate terms of equal power in $\lambda$, the resulting equations
are
\begin{equation}
\Biggl[\frac{\partial}{\partial\tau}+(-\Delta_{x}+m^{2}_{0}\,)
\Bigg]\varphi^{(0)}(\tau,x)=\eta(\tau,x), \label{42}
\end{equation}
\begin{equation}
\Biggl[\frac{\partial}{\partial\tau}+(-\Delta_{x}+m^{2}_{0}\,)
\Bigg]\varphi^{(1)}(\tau,x)=-\frac{1}{3!}
\left(\varphi^{(0)}(\tau,x)\right)^{3}, \label{43}
\end{equation}
and so on. Using the retarded Green function and assuming that
$\varphi^{\,(q)}(\tau,x)|_{\tau=0}=0,\,\,\forall\,q$, the solution
to the first equation given by Eq.(\ref{42}) can be written formally
as
\begin{equation}
\varphi^{(0)}(\tau,x)=\int_{0}^{\tau}d\tau'\int_{\Omega}d^{d}x'\,
G(\tau-\tau',x-x')\eta(\tau',x'). \label{sol}
\end{equation}
The second equation given by Eq.(\ref{43}) can also be solved using
the above result. We obtain
\begin{eqnarray}
\varphi^{(1)}(\tau,x)&=&
-\frac{1}{3!}\int_{0}^{\tau}d\tau_{1}\int_{\Omega}d^{d}x_{1}\,
G(\tau-\tau_{1},x-x_{1})\nonumber \\
&&\left(\int_{0}^{\tau_{1}}d\tau'\int_{\Omega}d^{d}x'\,
G(\tau_{1}-\tau',x_{1}-x')\eta(\tau',x') \right)^{3}. \label{44}
\end{eqnarray}
We have seen that we can generate all the tree diagrams with the
noise field contributions. We can also consider the $n$-point
correlation function
$\langle\,\varphi(\tau_{1},x_{1})\varphi(\tau_{2},x_{2})...
\varphi(\tau_{n},x_{n})\,\rangle_{\eta}$. Substituting the above
results in the $n$-point correlation function, and taking the random
averages over the white noise field using the Wick-decomposition
property defined by Eq.(\ref{388}) we generate the stochastic
diagrams. Each of these stochastic diagrams has the form of a
Feynman diagram, apart from the fact that we have to take into
account that we are joining together two white random noise fields
many times. Besides, the rules to obtain the algebraic values of the
stochastic diagrams are similar to the usual Feynman rules.

As simple examples let us show how to derive the two-point function
in the zeroth order $\langle\,\varphi(\tau_{1},x_{1})
\varphi(\tau_{2},x_{2})\,\rangle^{(0)}_{\eta}$, and also the first
order correction to the scalar two-point-function given by
$\langle\,\varphi(\tau_{1},x_{1})
\varphi(\tau_{2},x_{2})\,\rangle^{(1)}_{\eta}$. Using the
Eq.(\ref{26}) and the Einstein relations we have
\begin{equation}
\langle\,\varphi(\tau_{1},x_{1})
\varphi(\tau_{2},x_{2})\,\rangle^{(0)}_{\eta} =2
\int_{0}^{min(\tau_{1},\tau_{2})}d\tau'\int_{\Omega}d^{d}x'\,
G(\tau_{1}-\tau',x_{1}-x')\,G(\tau_{2}-\tau',x_{2}-x').
 \label{inc1}
\end{equation}
For the first order correction we get:
\begin{eqnarray}
&&\langle\,\varphi(X_{1})
\varphi(X_{2})\,\rangle^{(1)}_{\eta}=\nonumber\\
&& = -\frac{\lambda}{3!}\langle
\int\,dX_{3}\int\,dX_{4}\Biggl(G(X_{1}-X_{4})G(X_{2}-X_{3})+
G(X_{1}-X_{3})G(X_{2}-X_{4})\Bigg)\nonumber\\
&& \eta(X_{3})
\Biggl(\int\,dX_{5}\,G(X_{4}-X_{5})\eta(X_{5})\Bigg)^{3}\rangle_{\eta}.
\label{inc2}
\end{eqnarray}
where, for simplicity, we have introduced a compact notation:
\begin{equation}
\int_{0}^{\tau}d\tau\int_{\Omega}d^{d}x\equiv\int\,dX,
\end{equation}
and also $\varphi(\tau,x)\equiv\varphi(X)$ and finally
$\eta(\tau,x)\equiv\eta(X)$.

The process can be repeated and therefore the stochastic
quantization can be used as an alternative approach to describe
scalar quantum fields. Therefore, the two-point function up to the
first order level in the coupling constant $\lambda$ is given by

\begin{equation}
\langle\,\varphi(\tau_{1},x_{1})
\varphi(\tau_{2},x_{2})\,\rangle^{(1)}_{\eta} = (a) + (b) + (c),
\end{equation}
where $(a)$ is the zero order two-point function and (b) and (c)
are given, respectively, by:
\begin{equation}
(b)=-\frac{\lambda}{2}\,\delta^d(k_1+k_2)\int\,d^dk\int_{0}^{\tau_{1}}\,d\tau\,
G(k_1;\tau_1-\tau)D(k;\tau,\tau)D(k_2;\tau_2,\tau), \label{1}
\end{equation}
\begin{equation}
(c)=
-\frac{\lambda}{2}\,\delta^d(k_1+k_2)\int\,d^dk\int_{0}^{\tau_{2}}\,d\tau\,
G(k_2;\tau_2-\tau)D(k;\tau,\tau)D(k_1;\tau_1,\tau).\\
\end{equation}
These are the contributions in first order. A simple computation
shows that we recover the correct equilibrium result at equal
asymptotic Markov parameters ($\tau_1=\tau_2\rightarrow \infty$):
\begin{equation}
(b)|_{\tau_1=\tau_2\rightarrow
\infty}=-\frac{\lambda}{2}\,\delta^d(k_1+k_2)\frac{1}{(k_2^2+m_0^2)}
\frac{1}{(k_1^2+k_2^2+2m_0^2)} \int\,d^dk\frac{1}{(k^2+m_0^2)}.
\label{700}
\end{equation}

Obtaining the Schwinger functions in the asymptotic limit does not
guarantee that we gain a finite physical theory. The next step is
to implement a suitable regularization scheme. A crucial point to
find a satisfactory regularization scheme is to use one that
preserves the symmetries of the original model. The presence of
the Markov parameter as an extra dimension lead us to a new
regularization scheme, the stochastic regularization method, which
preserves all the symmetries of the theory under study. Therefore,
let us implement a continuum regularization procedure \cite{halp}.
We begin with a regularized Markovian Parisi-Wu Langevin system:
\begin{equation}
\frac{\partial}{\partial\tau}\varphi(\tau,x)
=-\frac{\delta\,S_0}{\delta\,\varphi(x)}|_{\varphi(x)=\varphi(\tau,\,x)}+
\int d^{d}y\,R_{xy}(\Delta)\, \eta(\tau,y). \label{200}
\end{equation}
The Einstein relations, given by Eqs. (\ref{28}) and (\ref{29}),
are maintained. The regulator $R(\Delta)$ that multiplies the
noise is a function of the Laplacian:
\begin{equation}
\Delta_{xy} = \int
d^{d}z\,(\partial_{\mu})_{xz}(\partial_{\mu})_{zy},\label{201}
\end{equation}
where
\begin{equation}
(\partial_{\mu})_{xy} = \partial_{\mu}^x \delta^d(x-y).\label{202}
\end{equation}
We will be working with a heat kernel regulator with the form:
\begin{equation}
R(\Delta;\Lambda) = \exp\biggl(\frac{\Delta}{\Lambda^2}\biggr),
\label{203}
\end{equation}
where $\Lambda$ is a parameter introduced to regularize the
theory. The basic restrictions on the form of this heat kernel
regulator are:
\begin{equation}
R(\Delta;\Lambda)|_{\,\Lambda\rightarrow\infty} = 1,
\end{equation}
or
\begin{equation}
R_{xy}(\Delta;\Lambda)|_{\,\Lambda\rightarrow\infty} =
\delta^d(x-y),
\end{equation}
which guarantees that the regularized process given by
Eq.(\ref{200}) reduces to the formal process given by
Eq.(\ref{23}) in the formal regulator limit
$\Lambda\rightarrow\infty$.

 With this modification in the Langevin
equation, it is possible to show that all the contributions to the
n-point function at all orders in the coupling constant $\lambda$
are finite. For instance, the contribution to the two-point
function at the one-loop level given by Eq.(\ref{1}) is rewritten
as:
\begin{equation}
(b)|_{\tau_1=\tau_2\rightarrow
\infty}=-\frac{\lambda}{2}\,\delta^d(k_1+k_2)\frac{R_{k_2}^2}{(k_2^2+m_0^2)}
\frac{1}{(k_1^2+k_2^2+2m_0^2)}
\int\,d^dk\frac{R_{k}^2}{(k^2+m_0^2)}, \label{204}
\end{equation}
where $R_k$ is the Fourier transform of the regulator, i.e.,
\begin{equation}
R_k(\Lambda) = R(\Delta;\Lambda)|_{\Delta = -k^2}.
\end{equation}

Now we use this method to discuss the quantization of scalar
theories with self-interaction in a curved spacetime without event
horizon and in a flat manifold with event horizon. Being more
specific, we are interested to investigate the
$\lambda\varphi^{4}$ theory in the Einstein and Rindler spacetime,
respectively.

\section{Stochastic quantization for the
$(\lambda\varphi^{4})_{d}$ scalar theory: the Einstein case}

\quad {\,\,} The aim of this section is to implement the stochastic
quantization and the stochastic regularization for the
self-interacting $\lambda\varphi^{4}$ theory in the one-loop level
in the Einstein spacetime. Let us consider a $M^{4}$ manifold that
admit a non-vanishing timelike Killing vector field $X$. If one can
always introduce coordinates $t=x^{0},x^{1},x^{2},x^{3}$ locally
such that $X=\frac{\partial}{\partial\,t}$ and the components of the
metric tensor are independent of $t$, $M^{4}$ is stationary. If
further the distribution $X^{\bot}$ of $3$-planes orthogonal to $X$
is integrable, then $M^{4}$ is static. Each integral curve of the
Killing vector vector field $X=\frac{\partial}{\partial\,t}$ is a
world line of an possible observer. Since
$X=\frac{\partial}{\partial\,t}$ generates isometries, the
$3$-planes $X^{\bot}$ are invariant under these isometries. For
static manifold, it is possible to perform a Wick rotation, i.e.,
analytically extend the pseudo-Riemannian  manifold to the
Riemannian domain without problem. Therefore for static spacetime
the implementation of the stochastic quantization is
straightforward.

In the previous section, we have been working in an Euclidean
space $R\,^{d}\times R\,^{(+)}$, where $R\,^{d}$ is the usual
Euclidean space and $R\,^{(+)}$ is the Markov sector. Now let us
generalize this to a more complicated case, i.e., let us work in a
general (and, for the time being, static) Riemannian manifold $M$.
In other words, we will consider a classical field theory defined
in a $M\times R\,^{(+)}$ manifold coupled with a heat reservoir.
Therefore, in a four-dimensional curved manifold, the Parisi-Wu
Langevin equation for scalar fields reads
\begin{equation}
\frac{\partial}{\partial\tau}\varphi(\tau,x)
=-\frac{1}{\sqrt{g}}\,\frac{\delta\,S_0}{\delta\,\varphi(x)}|_{\varphi(x)=\varphi(\tau,\,x)}+
\eta(\tau,x), \label{208}
\end{equation}
where $g = \det g_{\mu\nu}$ and the classical Euclidean action
$S_{0}$ is given by
\begin{equation}
S_0 = \int\, d^{4}x\,\sqrt{g}\,{\cal{L}}.
 \label{209}
\end{equation}
In the above equation $\cal{L}$ is given by
\begin{equation}
{\cal{L}}=
\frac{1}{2}\,g_{\mu\nu}\,\partial_{\mu}\varphi\partial_{\nu}\varphi
+ \frac{1}{2}\,(m^2 + \xi R)\varphi^{2}. \label{210}
\end{equation}
Note that we introduce a coupling between the scalar field and the
gravitational field represented by the term $\xi R\varphi^{2}$,
where $\xi$ is a numerical factor and $R$ is the Ricci scalar
curvature. The random noise field $\eta(\tau,x)$ obeys the
following generalized Einstein relations:
\begin{equation}
\langle\,\eta(\tau,x)\,\rangle_{\eta}=0, \label{211}
\end{equation}
and
\begin{equation}
\langle\, \eta(\tau,x)\,\eta(\tau',x')\,
\rangle_{\eta}\,=\,\frac{2}{\sqrt{g(x)}}\,\delta^{4}(x-x')\,\delta(\tau-\tau').
\label{212}
\end{equation}
Substituting Eq.(\ref{209}) and Eq.(\ref{210}) in the Langevin
equation given by Eq.(\ref{208}), we get:
\begin{equation}
\frac{\partial}{\partial\tau}\varphi(\tau,x) =-\biggl(-\Delta +
m^2 + \xi R\biggr)\varphi(\tau,x) + \eta(\tau,x), \label{213}
\end{equation}
where $\Delta$ is the four-dimensional Laplace-Beltrami operator
defined by:
\begin{eqnarray}
\Delta &=& g^{-1/2}\partial_{\mu}(g^{1/2}g_{\mu\nu}\partial_{\nu})
\nonumber\\
&=&g_{\mu\nu}\nabla_{\mu}\nabla_{\nu},
 \label{214}
\end{eqnarray}
$\nabla$ denoting the covariant derivative.

To proceed, as in the flat situation, let us introduce the
retarded Green function for the diffusion problem
$G(\tau-\tau',x,x')$, which obeys:
\begin{equation}
\Biggl[\frac{\partial}{\partial\tau}+\biggl(-\Delta_{x} + m^2 +
\xi R(x)\,\biggr) \Bigg]G(\tau-\tau',x,x')=
\frac{1}{\sqrt{g}}\,\delta^{4}(x-x')\delta(\tau-\tau'),
\label{215}
\end{equation}
if $\tau-\tau'>0$, and $G(\tau-\tau',x,x')=0$ if $\tau-\tau'<0$.

Using the retarded Green function and the initial condition
$\varphi(\tau,x)|_{\tau=0}=0$, a formal solution to Eq.(\ref{213})
reads
\begin{equation}
\varphi(\tau,x)=\int_{0}^{\tau}d\tau'\int_{\Omega}d^{4}x'\,\sqrt{g(x')}\,
G(\tau-\tau',x-x')\eta(\tau',x'). \label{2666}
\end{equation}
We see that, in order to solve Eq.(\ref{213}) we need to invert
the differential operator given by Eq.(\ref{214}). Depending on
the metric $g_{\mu\nu}$ defined on the manifold, this could be a
difficult task. In the Euclidean case this is easily solved, since
we decompose the fields in Fourier modes (plane waves), which are
solution to the Klein-Gordon equation in a flat metric; each
Fourier mode, as we have seen, obeys a Langevin equation. So, if
the Klein-Gordon equation in a general curved manifold allows us
to obtain solutions which can be decomposed in modes separated in
a similar manner to the flat case, we may as well decompose our
fields in these modes in a way that each of them, again, will obey
a Langevin equation. Then, Eq.(\ref{213}) can be solved in a
simple and direct way. Of course, this procedure will help us if
we admit manifolds with global Killing vectors, which,
unfortunately, is not always the case. So admitting that this is
the case, we may write the mode decompositions as:
\begin{equation}
\varphi(\tau,x) = \int\,d\tilde{\mu}(k)\varphi_{k}(\tau)u_{k}(x),
\label{216}
\end{equation}
and
\begin{equation}
\eta(\tau,x) = \int\,d\tilde{\mu}(k)\eta_{k}(\tau)u_{k}(x),
\label{217}
\end{equation}
where the measure $\tilde{\mu}(k)$ depends on the metric we are
interested in. For instance, in the flat case, we have that in a
four dimensional space $d\tilde{\mu}(k) = d^{4}k$ and the modes
$u_{k}(x)$ are given by:
\begin{equation}
u_{k}(x) = \frac{1}{(2\pi)^2} e^{ikx}.
\end{equation}
So we see that, in the flat case, Eqs. (\ref{216}) and (\ref{217})
reduce to the Fourier decomposition given by Eqs.(\ref{33}) and
(\ref{34}), respectively.

Now let us apply this formalism to the Euclidean Einstein
manifold. The general Robertson-Walker line element is given by
\cite{davies} \cite{grib}:
\begin{equation}
ds^{2}= - dt^{2} + a^2(t)\sum_{i,j=1}^3 h_{ij}dx^{i}dx^{j},
\label{205}
\end{equation}
where:
\begin{eqnarray}
\sum_{i,j=1}^3 h_{ij}dx^{i}dx^{j} &=&
\left(1-Kr^2\right)^{-1}dr^{2}+r^{2}(d\theta^{2} + \sin^2\phi
d\phi^{2}) \nonumber\\
&&= d\chi^{2} + f^2(\chi)(d\theta^{2} + \sin^2\phi d\phi^{2}),
\label{206}
\end{eqnarray}
and, for the Einstein manifold we have ($K = 1$), $f(\chi) = r =
\sin\chi$, $0\leq \chi \leq 2\pi$.

The Eq.(\ref{206}) gives the line element on the spatial sections
in the pseudo-Riemannian case which are hyperbolic, flat or closed
depending on whether $K = -1, 0, 1$, respectively. Writing
$C(\rho) = a^2(t)$, with the conformal time parameter $\rho$ given
by
\begin{equation}
\rho = \int^{t}\, dt'a^{-1}(t'),
\end{equation}
the line element can be recast in the form:
\begin{equation}
ds^{2}= C(\rho)\biggl(-d\rho^{2}+\sum_{i,j=1}^3
h_{ij}dx^{i}dx^{j}\biggr). \label{207}
\end{equation}
In the simplest case, namely static space-times with $C(\rho)= C =
a^2 =$ constant, the scalar curvature is given by $R =
\frac{6K}{C}$. After the Wick rotation, we are working with the
Euclidean Einstein manifold. In this case, we may decompose the
modes as:
\begin{equation}
u_{k}(x) = C^{-1/2}\textbf{X}_{\vec{k}}(\vec{x})e^{ik_{\rho}\rho},
\label{218}
\end{equation}
with $\vec{x} = (r,\theta,\phi)$ or $(\chi,\theta,\phi)$ and
$\textbf{X}_{\vec{k}}$ being a solution of
\begin{equation}
\triangle^{(3)} \textbf{X}_{\vec{k}} = -(q^{2} -
K)\,\textbf{X}_{\vec{k}}. \label{219}
\end{equation}
In Eq.(\ref{219}), $\triangle^{(3)}$ is the Laplacian associated
with the spatial metric $h_{ij}$:
\begin{equation}
\Delta = h^{-1/2}\partial_{i}(h^{1/2}h^{ij}\partial_{j}).
\label{220}
\end{equation}
Further, the functions $\textbf{X}_{\vec{k}}$ are normalized such
that:
\begin{equation}
\int\,d^{3}x\,h^{\frac{1}{2}}\,\textbf{X}_{\vec{k}}
(\vec{x})\textbf{X}_{\vec{k'}}^{*}(\vec{x})
= \delta(\vec{k},\vec{k'}), \label{221}
\end{equation}
where $\delta(\vec{k},\vec{k'})$ is the $\delta$-function with
respect to the measure $\tilde{\mu}$:
\begin{equation}
\int\,d\tilde{\mu}(k')\,f(\vec{k})\delta(\vec{k},\vec{k'}) =
f(\vec{k'}). \label{222}
\end{equation}
The eigenfunctions $\textbf{X}_{\vec{k}}$ of the three-dimensional
Laplacian are, for $K=1$:
\begin{equation}
\textbf{X}_{\vec{k}}(\vec{x}) =
\Pi_{qJ}^{+}(\chi)Y_{MJ}(\theta,\phi),\label{223}
\end{equation}
with $\vec{k} = (q,J,M)$, $M = -J, -J+1, ... , J$, $J = 0,1, ... ,
q-1$ and $q = 1,2, ...$. The $Y_{MJ}$ are the usual spherical
harmonics. The functions $\Pi_{qJ}^{+}$ can be defined from
\cite{bander}:
\begin{equation}
\Pi_{qJ}^{-}(\chi) = \biggl[\frac{1}{2}\pi q^{2}(q^{2}+ 1) ...
(q^{2} + J^{2})\biggr]^{-1/2}\sinh^{J}\chi\biggl(\frac{d}{d\cosh
\chi}\biggr)^{J+1}\cos q\chi, \label{224}
\end{equation}
by replacing $q$ by $-iq$ and $\chi$ by $-i\chi$ in the latter.

With these definitions, the measure $\tilde{\mu}(k)$ for the
Einstein universe is defined as follows:
\begin{equation}
\int\,d\tilde{\mu}(k) = \frac{1}{2\pi}\int\,dk_{\rho}\sum_{q, J,
M}. \label{225}
\end{equation}
So, inserting the mode decomposition given by Eq.(\ref{218}) in
Eq.(\ref{213}) we have that each mode coefficient satisfy the
Langevin equation given by
\begin{equation}
\frac{\partial}{\partial\tau}\varphi_{k}(\tau)
=-\frac{1}{C}(k^{2}+\mu^{2})\varphi_{k}(\tau)+ \eta_{k}(\tau),
\label{226}
\end{equation}
where $k^{2} = q^{2} + k_{\rho}^{2}$ and $\mu^{2} = Cm^{2} +
(6\xi-1)K$ and $K = 1$. For simplicity, we redefine
$\frac{1}{C}(k^{2}+\mu^{2})\rightarrow(k^{2}+\mu^{2})$.

The solution for Eq.(\ref{226}), with the initial condition
$\varphi_{k}(\tau)|_{\tau=0}=0$, reads:
\begin{equation}
\varphi_{k}(\tau)= \int_{0}^{\tau}d\tau'G_{k}(\tau,\tau')
\eta(\tau',k), \label{227}
\end{equation}
where
\begin{equation}
G_{k}(\tau,\tau')= \exp\left(-(k^{2}+ \mu^{2})(\tau-\tau')\right)
\theta(\tau-\tau') \label{228}
\end{equation}
is the retarded Green function for the diffusion problem.

Using the Einstein relations, we get that the mode coefficients for
the random noise satisfies
\begin{equation}
\langle\,\eta_{k}(\tau)\,\rangle_{\eta}=0 \label{229}
\end{equation}
and
\begin{equation}
\langle\,\eta_{k}(\tau)\eta_{k'}(\tau')\,\rangle
_{\eta}=2\delta(\tau-\tau')\delta^{4}(k,k'), \label{230}
\end{equation}
where $\delta^{4}(k,k')=\delta(k_{\rho} +
k'_{\rho})\delta_{qq'}\delta_{JJ'}\delta_{MM'}$.\\
The two-point function $D_{k}(\tau,\tau')$ can be calculated in a
similar way as in the Euclidean flat case. We have
\begin{equation}
D_{k}(\tau,\tau')=\frac{1}{(2\pi)}\delta^{4}(k,k')\frac{1}{(k^{2}+
\mu^{2})}\left(e^{-((k^{2}+\mu^{2})|\tau - \tau'|)}-
e^{-((k^{2}+\mu^{2})(\tau + \tau'))}\right), \label{231}
\end{equation}
or, in the ``coordinate" space:
\begin{eqnarray}
D(\tau,\tau';x,x')&=&
\int\,d\tilde{\mu}(k)u_{k}(x)u_{k}^{*}(x')D_{k}(\tau,\tau')
= \nonumber \\
&&\int\,d\tilde{\mu}(k)u_{k}(x)u_{k}^{*}(x')\frac{1}{(k^{2}+
\mu^{2})}\left(e^{-((k^{2}+\mu^{2})|\tau - \tau'|)}-
e^{-((k^{2}+\mu^{2})(\tau + \tau'))}\right), \label{232}
\end{eqnarray}
where $\int\,d\tilde{\mu}(k)$ is given by Eq.(\ref{225}).

Now, let us apply this method for the case of a self-interacting
theory with an interaction action given by:
\begin{equation}
S_{I}[\varphi]= \int d^{4}x\,\sqrt{g(x)}\frac{\lambda}{4!}
\,\varphi^{4}(x). \label{233}
\end{equation}
In the same way, we can solve the equation using a perturbative
series in $\lambda$. The two-point function up to the one loop
level is given by:
\begin{equation}
\langle\,\varphi(\tau_{1},x_{1})
\varphi(\tau_{2},x_{2})\,\rangle^{(1)}_{\eta} = (a) + (b) + (c),
\label{234}
\end{equation}
where $(a)$ is the zero order two-point function given by
Eq.(\ref{231}) and (b) and (c) are given respectively by
\begin{equation}
(b)=-\frac{\lambda}{2}\,\delta^4(k_1,k_2)\int\,d\tilde{\beta}(k)\int_{0}^{\tau_{1}}\,d\tau\,
G_{k_1}(\tau_1-\tau)D_{k}(\tau,\tau)D_{k_2}(\tau_2,\tau),
\label{235}
\end{equation}
\begin{equation}
(c)=
-\frac{\lambda}{2}\,\delta^4(k_1,k_2)\int\,d\tilde{\beta}(k)\int_{0}^{\tau_{2}}\,d\tau\,
G_{k_2}(\tau_2-\tau)D_{k}(\tau,\tau)D_{k_1}(\tau_1,\tau),
\label{236}
\end{equation}
where $\int\,d\tilde{\beta}(q)
\equiv\frac{1}{2\pi}\int\,dk_{\rho}\sum_{q, J, M}
\textbf{X}_{\vec{q}}\textbf{X}_{\vec{q}}$. These are the
contributions in first order. A simple computation shows that we
obtain, at equal asymptotic Markov parameters
($\tau_1=\tau_2\rightarrow \infty$):
\begin{equation}
(b)|_{\tau_1=\tau_2\rightarrow
\infty}=-\frac{\lambda}{2}\,\delta^4(k_1,k_2)\frac{1}{(k_2^2+\mu^2)}
\frac{1}{(k_1^2+k_2^2+2\mu^2)}
\int\,d\tilde{\beta}(k)\frac{1}{(k^2+\mu^2)}. \label{237}
\end{equation}
\\
Now we define the quantity $I$ as:
\begin{eqnarray}
I &=& \int\, d\tilde{\beta}(q)\frac{1}{(k^2+\mu^2)} \nonumber \\
&& = \frac{1}{2\pi}\int\,dk_{\rho}\sum_{q, J,
M}\textbf{X}_{\vec{q}}\textbf{X}_{\vec{q}}\frac{1}{(k^2+\mu^2)} \nonumber \\
&& =
\int\,dk_{\rho}\sum_{q}\textbf{X}_{\vec{q}}\textbf{X}_{\vec{q}}\frac{q^2}{(q^2+b^2)},
\label{238}
\end{eqnarray}
where $b^2 = k_{\rho}^2 + \mu^2$. It is easy to show that the
series in this equation is divergent. So, we need a procedure to
regularize it and obtain a finite quantity for the two-point
function. As we will see now, this can be done within the
covariant stochastic regularization \cite{halp}.

The generalization of Eq.(\ref{200}) to general four-dimensional
spacetimes is
\begin{equation}
\frac{\partial}{\partial\tau}\varphi(\tau,x)
=-\frac{1}{\sqrt{g}}\,\frac{\delta\,S_0}{\delta\,\varphi(x)}|_{\varphi(x)=\varphi(\tau,\,x)}+
\int d^{4}y\,\sqrt{g}\,R_{xy}(\Delta)\, \eta(\tau,y), \label{239}
\end{equation}
where, for covariant reasons, the regulator is now a function of
the covariant Laplacian:
\begin{equation}
\Delta_{xy} = \int
d^{4}z\,(\nabla_{\mu})_{xz}(\nabla_{\mu})_{zy},\label{2011}
\end{equation}
with
\begin{equation}
(\nabla_{\mu})_{xy} = \nabla_{\mu}^x \delta^4(x-y).\label{2021}
\end{equation}

Using the mode decomposition given by Eq.(\ref{218}), the
Eq.(\ref{239}) reduces to
\begin{equation}
\frac{\partial}{\partial\tau}\varphi_{k}(\tau)
=-\frac{1}{C}(k^{2}+\mu^{2})\varphi_{k}(\tau)+ \eta_{k}(\tau)R_{k},
\label{240}
\end{equation}
where $R_{k} = R_{xy}(\Delta)|_{\Delta = -k^2 + K} $ and
\begin{equation}
 R_{xy}(\Delta)=
\int\,d\tilde{\mu}(k)u_{k}(x)u_{k}^{*}(y)R_{k}. \label{241}
\end{equation}
Then, the solution for the regularized Langevin equation follows:
\begin{equation}
\varphi_{k}(\tau)= \int_{0}^{\tau}d\tau'G_{k}(\tau,\tau')
\eta(\tau',k), \label{242}
\end{equation}
where $G_{k}(\tau,\tau')$ is given by Eq.(\ref{228}). Then, it is
easy to show that the zero order two-point function and the
regularized contribution $(b)_{\Lambda}$ at the first order at the
coupling constant are given by, respectively:
\begin{equation}
D_{k}(\tau,\tau')=\frac{1}{(2\pi)}\delta^{4}(k,k')\frac{R_{k}^2}{(k^{2}+
\mu^{2})}\left(e^{-((k^{2}+\mu^{2})|\tau - \tau'|)}-
e^{-((k^{2}+\mu^{2})(\tau + \tau'))}\right), \label{243}
\end{equation}
and
\begin{equation}
(b)_{\Lambda}|_{\tau_1=\tau_2\rightarrow
\infty}=-\frac{\lambda}{2}\,\delta^4(k_1,k_2)\frac{R_{k_2}^2}{(k_2^2+\mu^2)}
\frac{1}{(k_1^2+k_2^2+2\mu^2)}
\int\,d\tilde{\beta}(k)\frac{R_{k}^2}{(k^2+\mu^2)}. \label{244}
\end{equation}
Let us isolate the part where we had problems. We have:
\begin{eqnarray}
I_{\Lambda} &=& \int\,d\tilde{\beta}(k)\frac{R_{k}^2}{(k^2+\mu^2)} \nonumber \\
&& = \frac{1}{2\pi}\int\,dk_{\rho}\sum_{q, J,
M}\textbf{X}_{\vec{q}}\textbf{X}_{\vec{q}}\,\,\frac{R_{k}^2}{(k^2+\mu^2)}.
\label{245}
\end{eqnarray}
After some tedious calculations \cite{prud} \cite{grad}
\cite{erdelyi}, we arrive at:
\begin{equation}
I_{\Lambda} = \frac{1}{4}\,e^{\frac{2K}{C\Lambda^2} +
\frac{2}{C\Lambda^2}\mu^2} \int_{\mu}^{\infty}\,dx
(x^2-\mu^{2})^{1/2} erfc(\sqrt{\alpha}x), \label{246}
\end{equation}
where $erfc(x)$ is the complementary error function. At the
expense of considerable labor, it is possible to extend our
results to the four-point function.

A quite important point is that this regularization procedure
preserves all the symmetries of the unregularized Lagrangian,
since it is not an action regularization. The next step would be
to isolate the parts that go to infinity in the limit
$\Lambda\rightarrow\infty$ and remove them with a suitable
redefinition of the constants of the theory, i.e., carry out the
renormalization program. A natural question now would be if we can
actually renormalize all the $n$-point functions at all orders at
the coupling constant $\lambda$. Birrel \cite{birrel} has given
arguments that a priori we cannot expect that a regularization
independent proof of the renormalizability of the
$\lambda\varphi^{4}$ theory in a curved background exists. One
attempt of general proof of renormalizability of
$\lambda\varphi^{4}$ theory defined in a spacetime which can be
analytically continued to Euclidean situation was given by Bunch
\cite{bunch}. Using the Epstein-Glaser method, Brunetti and
Fredenhagen \cite{bru} presented a perturbative construction of
this theory on a smooth globally hyperbolic curved spacetime.

Our derivation shows that the stochastic regularization may be an
attempt in a direction of such regularization independent proof,
even though we are still restricted to the same situation studied by
Bunch. Indeed, we know that, in this case of the Einstein universe,
the free two-point function presents itself as a sum of Minkowski
space positive frequency two-point functions, so it is natural to
expect that, as it is in the Minkowski case, a regularization
independent proof of the renormalizability of such interacting
theory does exist, at least for the Einstein universe.

\section{Stochastic quantization for the
$(\lambda\varphi^{4})_{d}$ scalar theory: the Rindler case}

\quad Now let us consider the stochastic quantization in a
spacetime with an event horizon. The Schwarzschild black hole is
the most familiar solution to the vacuum Einstein equations, which
has an event horizon. At the origin, there is a curvature
singularity and at $R=2M$ we have a singularity due to the bad
behavior of this particular coordinate system. Although it is not
possible to extend the spacetime across the singularity, it is
possible to analytically continue the manifold from $r>2M$ to the
region $r<2M$, $(r\neq 0)$. The maximal extension of the manifold
described by the Schwarzschild line element with $2M<r<\infty$ is
the usual Kruskal extension \cite{kru}. The Kruskal spacetime
defines two outer asymptotically flat regions and also two regions
inside the event horizon, bounded by the past and future
singularities (see Appendix A). Since we know that, close to the
horizon, the Schwarzschild coordinates $t$ and $r$ behaves as
Rindler's spacetime coordinates, we shall investigate the
stochastic quantization in the Rindler metric. The background
material relevant for us can be found in the text book \cite{wald}
and also in the review articles \cite{strominger} \cite{brout}
\cite{ford} \cite{padmanabhan}.

Let us consider a d-dimensional flat Minkowski spacetime where we
are using the usual cartesian coordinates
$y^{\mu}=(y^{0},y^{1},...,y^{d-1})$. It is possible to find a
curvilinear coordinate system $x^{\mu}=(t,x^{1},...,x^{d-2},z)$,
called the Rindler's coordinate system and can be shown that this
coordinate system is one naturaly adapted to an observer with
constant proper acceleration. It is important to stress that this
coordinate system with the respective coordinate transformation
cover only the region $y^{d-1}>|y^{0}|$. Since this coordinate
system does not cover the whole Minkowski spacetime, we can define
three coordinate transformations with the respective coordinate
systems defined in different regions of the Minkowski spacetime.
These regions are known in the literature as Rindler's L, Milne F,
and finally Milne P. The four coordinate transformations and the
coordinate systems together cover all the Minkowski spacetime. The
coordinate system that cover the region inside the light cone is
the $d$-dimensional Milne spacetime.

The implementation of the canonical quantization in Rindler's
spacetime is very simple, since if the spacetime has a stationary
geometry, there is a time-like killing vector field $K$ that
generates one-parametric Lie group of isometries and the
orthogonal modes which satisfies
${\cal{L}}_{K}u_{n}(x)=-i\omega\,u_{n}$, where ${\cal{L}}_{K}$ is
the Lie derivative with respect to $K$ and the $u_{n}(x)$ are the
positive frequency modes. In this situation, there is a natural
way to define positive and negative frequency modes. Note that the
Rindler's line element is $t$ independent, and consequently there
is a straightforward way to define positive and negative frequency
modes in order to impose the canonical quantization in Rindler's
spacetime.

It has long be recognized that the Rindler's vacuum
$|\,0,R\rangle$ that appear in the Fulling canonical quantization
of a scalar field in the Rindler's spacetime is not unitarily
equivalent to the Minkowski vacuum $|\,0,M>$. Let us define the
following generating functionals, $Z(h)$ and $Z_{R}(h)$, i.e.,
\begin{equation}
Z(h)=\frac{\langle\,0,M,out|\,0,M,in\rangle_{h}}
{\left<0,M,out|\,0,M,in\right>_{h=0}} \label{17}
\end{equation}
and
\begin{equation}
Z_{R}(h)=\frac{\left<0,R,out|\,0,R, in\right>_{h}}
{\left<0,R,out|\,0,R,in\right>_{h=0}} \label{18}
\end{equation}
where $|\,0,M,in\rangle$, $|\,0,M,out\rangle$, $|\,0,R,in\rangle$
and finally $|\,0,R,out\rangle$ are the IN and OUT vacuum states
for the theory with the Minkowski and Rindler Hamiltonian density
respectively.

Starting from $Z(h)$ it is possible after analytic continuation
and imposing periodicity in the Euclidean time, to define
$Z(\beta;h)$, i.e., the finite temperature Schwinger functional.
It is clear that
$lim_{\beta\rightarrow\infty}Z(\beta;h)=Z(\infty;h)\equiv Z(h)$,
where $Z(\infty;h)$ is the zero temperature Euclidean functional
which generates the Schwinger functions in the whole Euclidean
space. At the same way it is possible to define $Z_{R}(\beta;h)$,
i.e. the finite temperature Euclidean Rindler functional. Again we
have
$lim_{\beta\rightarrow\infty}Z_{R}(\beta;h)=Z_{R}(\infty;h)\equiv
Z(h)$, where $Z_{R}(\infty;h)$ is the zero temperature Euclidean
functional which generates the Schwinger functions in the analytic
extended Rindler manifold.

From the above definitions, let us define the following Schwinger
functions. We are following the discussion of Christensen and Duff
\cite{cd}:

i) $G_{\beta}(x,x')$ two-point function, i.e. the Schwinger
function obtained from the finite temperature Schwinger
functional.

ii) $G_{\infty}(x,x')$ two-point function, i.e. the Schwinger
function obtained from the zero temperature Schwinger functional.

Now, let us analyse the two-point function that we are able to
construct in the Euclidean manifold. Remember that the Rindler
Euclidean metric possesses a coordinate singularity at the origin
where $z=0$. The point $z=0$ is a conical point of the Euclidean
manifold but is a regular point if the Euclidean time is periodic.

i) $G^{(1)}(x,x')$ two-point function which is periodic in the
Euclidean Rindler's time. This is the usual scalar two-point
Schwinger function. It is a function of the geodesic distance
beween two points in the manifold. Since all path around the
origin are topologically equivalent, therefore  $G^{(1)}(x,x')$
should be periodic in Euclidean Rindler's time. Therefore these
Green's functions can be generated by functional derivatives in
the $Z(\beta;h)$, i.e., the finite temperature Schwinger
functional.

ii) $G^{(0)}(x,x')$ two-point function. This two-point function
appears in the situation where the Euclidean space has a hole in
the origin. In the case where the two points are on the circular
world-line it is possible to relate the two two-point functions
$G^{(1)}(x,x')$ and $G^{(0)}(x,x')$. This can be done using a
simple trigonometric identity
\begin{equation}
G^{(1)}(\tau,\tau')=\sum_{n=-\infty}^{\infty}G^{0}(\tau,\tau'+\frac{2\pi
n}{a}). \label{18}
\end{equation}
Since the finite temperature Schwinger function must be periodic
in the Euclidean time with periodicity $\beta$, making the
identification $\beta=\frac{2\pi}{a}$ it is clear that
$G^{(1)}(x,x')= G_{\beta}(x,x')$. Also we have  $G^{(0)}(x,x')=
G_{\infty}(x,x')$. Therefore these Green's functions can be
generated by functional derivatives in the $Z(\infty;h)$ which is
the zero temperature Euclidean functional which generates the
Schwinger functions in the whole Euclidean space.

Now, after a Wick rotation, we should apply the stochastic
quantization for the four-dimensional Rindler space, which becomes
a multiple connected manifold, with the Euclidean metric:
\begin{equation}
ds{^2} = z^{2}dt^{2} + dx_{1}^{2} + dx_{2}^{2} + dz^{2}.
 \label{r1}
\end{equation}
In the Euclidean Rindler space, the Langevin equation for the
field $\varphi(x)$ reads:
\begin{equation}
\frac{\partial}{\partial\tau}\varphi(\tau,x)
=-\sqrt{g}\,\frac{\delta\,S_0}{\delta\,\varphi(x)}|_{\varphi(x)=\varphi(\tau,\,x)}+
\eta(\tau,x), \label{251}
\end{equation}
where $S_0$ is the Euclidean action for the scalar free fields.
Notice that this expression is quite different from the Langevin
equation for the Einstein metric, Eq.(\ref{208}), since we have
inverse power of the determinant of the metric multiplied by
$\frac{\delta\,S_0}{\delta\,\varphi(x)}$. So, with the Rindler
metric given above, we have:
\begin{equation}
\frac{\partial}{\partial\tau}\varphi(\tau,x) =-(-\Delta +
z^{2}m^{2}) \varphi(\tau,x) + \eta(\tau,x), \label{248}
\end{equation}
where the operator $\Delta$ is defined by:
\begin{equation}
\Delta =
\partial_{t}^{2} + z^{2}(\partial_{x_{1}}^{2}+\partial_{x_{2}}^{2}+\partial_{z}^{2})
+ z\partial_{z}.
 \label{250}
\end{equation}
To proceed with the implementation of the stochastic quantization,
we have to use the generalized Einstein relations, given by
Eq.(\ref{211}) and Eq.(\ref{212}).

From the modes presented by references \cite{candelas}
\cite{unruh}, we can obtain the following Euclidean modes:

\begin{equation}
u_{k\nu}(\vec{x},z) = \frac{1}{2\pi^2}\biggl(\nu
\sinh(\pi\nu)\biggr)^{1/2}e^{i\vec{k}.\vec{x}}K_{i\nu}(\mu z),
\label{252}
\end{equation}
where $\vec{k}.\vec{x}= k_{0}t + k_{1}x_{1} + k_{2}x_{2}$, $\mu =
\sqrt{\vec{k}^{2}+m^{2}}$ and $K_{\mu}(x)$ is the Macdonald
function. This are normalized such as:
\begin{equation}
\frac{1}{\pi^2}\int\,\frac{dz}{z}K_{i\nu}(\mu z)K_{i\nu'}(\mu z)=
\frac{\delta(\nu,\nu')}{(\nu+\nu')(\sinh(\pi\nu)\sinh(\pi\nu'))^{1/2}}.
\end{equation}
Now, we use the general mode decomposition, given by
Eq.(\ref{216}) and Eq.(\ref{217}), with the following measure:
\begin{equation}
\int\,d\tilde{\mu}(k) = \int_{0}^{\infty}\,d\nu\int\,d\vec{k} =
\int_{0}^{\infty}\,d\nu\int\,dk_1\,\int\,dk_2\int\,\frac{dk_{0}}{2\pi}.
\label{253}
\end{equation}
Remembering Eq.(\ref{252}) for the modes, we will have again that
each mode coefficient obeys a Langevin equation of the form:
\begin{equation}
\frac{\partial}{\partial\tau}\varphi_{k\nu}(\tau)
=-(k_{0}^{2}+\nu^{2})\varphi_{k\nu}(\tau)+ \eta_{k\nu}(\tau).
\label{254}
\end{equation}
The solution for Eq.(\ref{254}), with the initial condition
$\varphi_{k\nu}(\tau)|_{\tau=0}=0$, reads:

\begin{equation}
\varphi_{k\nu}(\tau)= \int_{0}^{\tau}d\tau'G_{k\nu}(\tau,\tau')
\eta(\tau',k), \label{255}
\end{equation}
where
\begin{equation}
G_{k\nu}(\tau,\tau')= \exp\left(-(k_{0}^{2}+
\nu^{2})(\tau-\tau')\right) \theta(\tau-\tau') \label{256}
\end{equation}
is the retarded Green function for the diffusion problem. Using the
Einstein relations, we get that the mode coefficients for the random
noise satisfies
\begin{equation}
\langle\,\eta_{k}(\tau)\,\rangle_{\eta}=0 \label{257}
\end{equation}
and
\begin{equation}
\langle\,\eta_{k}(\tau)\eta_{k'}(\tau')\,\rangle
_{\eta}=2\delta(\tau-\tau')\delta(k,k')\delta(\nu,\nu'), \label{258}
\end{equation}
where $\delta(k,k')=\delta(k_{0} + k'_{0})\delta(\vec{k} +
\vec{k'})$.

The two-point function $D^{0}_{k}(\tau,\tau')$ can be calculated
in a similar way as in the Einstein case. We have
\begin{equation}
D^{0}_{k\nu}(\tau,\tau')=\delta(k,k')\delta(\nu,\nu')\frac{1}{(k_{0}^{2}+
\nu^{2})}\left(e^{-((k_{0}^{2}+\nu^{2})|\tau - \tau'|)}-
e^{-((k_{0}^{2}+\nu^{2})(\tau + \tau'))}\right), \label{259}
\end{equation}
or, in the ``coordinate" space:
\begin{eqnarray}
D^{0}(x,x';\tau,\tau')&=&\frac{1}{4\pi^{4}}\int_{0}^{\infty}\,d\nu\int\,dk\,
\nu\sinh(\pi\nu)K_{i\nu}(\mu z)K_{i\nu}(\mu z')
\nonumber\\
&&\frac{e^{i\vec{k}.(\vec{x}-\vec{x'})}}{k_{0}^{2}+\nu^{2}}\left(e^{-((k_{0}^{2}+\nu^{2})|\tau
- \tau'|)}- e^{-((k_{0}^{2}+\nu^{2})(\tau + \tau'))}\right).
\label{260}
\end{eqnarray}
In the limit $\tau=\tau'\rightarrow \infty$, and proceeding with
similar steps as in reference \cite{candelas}, one may prove that
we have obtained the usual result for the free two-point function,
i.e., the Schwinger two-point function for a (straight-line)
geodesic distance between the points $x$ e $x'$. As is well known
\cite{cd}, this function is determined uniquely by the requirement
that, in the absence of any holes in space, one identifies the
``angle" $t$ (which is the Euclidean ``time") with the angle $t +
2\pi n$ ($n=$integer) because all paths around the origin are
topologically equivalent. Thus, as expected, $D^{0}$ should be
periodic in $t$ with period $2\pi$. But we know that, when we do
the analytical continuation $t\rightarrow it$ in the Rindler
metric, the event horizon, which in Minkwoski space is represented
as an impenetrable barrier to the accelerating observer,
translates in Euclidean language into the statement that closed
paths around the origin cannot be continuously shrunk to a point.
An inertial observer sees a space with the usual topology, $R^4$,
but the topology seen by the accelerating observer is different,
i.e., his space has a hole in it. Then, paths winding around the
origin cannot all be deformed into each other; they fall into
topologically distinct classes labelled by the winding number $n$.
Now, we do not identify $t$ with $t + 2\pi n$. This two-point
function, which we shall call $D^{1}$, exhibits an infinite
periodicity. To obtain it, we should work with the following modes
\cite{full}:
\begin{equation}
u_{k\nu}(\vec{x},z) = \frac{1}{2\pi^2}(\nu
\sinh(\pi\nu))^{1/2}e^{i(k_1x_1 + k_2x_2) - k_{0}\mid t
\mid}K_{i\nu}(\mu z), \label{252e}
\end{equation}
With this prescription, we obtain the same results as found in
literature \cite{zarro}:
\begin{eqnarray}
D^{0}(x,x';\tau,\tau') &=& \sum_{n=-\infty}^{n=+\infty}D^{1}(x,x';\tau,\tau')\nonumber\\
&& =
\frac{1}{4\pi^{4}}\sum_{n=-\infty}^{n=+\infty}\int_{0}^{\infty}\,d\nu\int\,dk\,
\nu\sinh(\pi\nu)K_{i\nu}(\mu z)K_{i\nu}(\mu z')
\nonumber\\
&&\frac{e^{i(k_1x_1 + k_2x_2)-k_{0}\mid t+2\pi
n\mid}}{k_{0}^{2}+\nu^{2}}\left(e^{-((k_{0}^{2}+\nu^{2})|\tau -
\tau'|)}- e^{-((k_{0}^{2}+\nu^{2})(\tau + \tau'))}\right),
\end{eqnarray}
where the geodesic distance between the points $x$ and $x'$ is an
arc length in the function $D^{1}$. It is interesting to note
that, using the modes given by Eq.(\ref{252e}) (see reference
\cite{full}), and then summing up all configurations with a
winding number $n$ we reach the same result as using the modes
given by Eq.(\ref{252}) (see reference \cite{candelas}).

Let us repeat the method for the case of self-interacting theory
with the interaction action given by:
\begin{equation}
S_{I}[\varphi]= \int d^{4}x\,\sqrt{g(x)}\frac{\lambda}{4!}
\,\varphi^{4}(x). \label{261}
\end{equation}
In this case, we have similar equations to the Einstein case.
Eq.(\ref{1}) reads, in the Rindler space:
\begin{equation}
(b)=-\frac{\lambda}{2}\,\delta(k_1,k_2)\delta(\nu_1,\nu_2)\int\,d\tilde{\beta}(k)\int_{0}^{\tau_{1}}\,d\tau\,
G_{k_1\nu_1}(\tau_1-\tau)D^{0}_{k\nu}(\tau,\tau)D^{0}_{k_2\nu_2}(\tau_2,\tau),
\label{262}
\end{equation}
which, in the limit $\tau=\tau'\rightarrow \infty$, becomes:
\begin{equation}
(b)|_{\tau_1=\tau_2\rightarrow
\infty}=-\frac{\lambda}{2}\,\delta(k_1,k_2)\delta(\nu_1,\nu_2)\frac{1}{(k_2^2+\nu_{2}^2)}
\frac{1}{(k_1^2+k_2^2+\nu_{1}^{2}+\nu_{2}^{2})}
\int\,d\tilde{\beta}(k)\frac{1}{(k^2+\nu^2)} \label{263}
\end{equation}
with similar divergences and
$\int\,d\tilde{\beta}\equiv\int\,d\nu\int\,dk\nu
\sinh(\pi\nu)K_{i\nu}(\mu z)K_{i\nu}(\mu z)$. Now, we may apply
the continuum regularization as before \cite{halp}, with a
regulator that is a function of the following operator:
\begin{equation}
\Delta' =
z^{2}(\partial_{x_1}^{2}+\partial_{x_2}^{2}+\partial_{z}^{2}) +
z\partial_{z} -z^{2}m^{2}.
 \label{264}
\end{equation}
The regularized Langevin equation reads:
\begin{equation}
\frac{\partial}{\partial\tau}\varphi_{k\nu}(\tau)
=-(k_{0}^{2}+\nu^{2})\varphi_{k\nu}(\tau)+
R_{k\nu}\eta_{k\nu}(\tau), \label{265}
\end{equation}
with the solution:
\begin{equation}
\varphi_{k\nu}(\tau)= \int_{0}^{\tau}d\tau'G_{k\nu}(\tau,\tau')
R_{k\nu}\eta(\tau',k). \label{266}
\end{equation}
where $R_{k\nu}(\Lambda) = R(\Delta';\Lambda)|_{\Delta' = \nu^{2}}
$. Then, it is easy to show that the zero order two-point function
and the regularized contribution $(b)_{\Lambda}$ at the one loop
level are given by, respectively:
\begin{equation}
D^{\Lambda}_{k\nu}(\tau,\tau')=\delta(k,k')\delta(\nu,\nu')\frac{R_{k\nu}^{2}}{(k_{0}^{2}+
\nu^{2})}\left(e^{-((k_{0}^{2}+\nu^{2})|\tau - \tau'|)}-
e^{-((k_{0}^{2}+\nu^{2})(\tau + \tau'))}\right), \label{267}
\end{equation}
and
\begin{equation}
(b)|_{\tau_1=\tau_2\rightarrow
\infty}=-\frac{\lambda}{2}\,\delta(k_1,k_2)\delta(\nu_1,\nu_2)\frac{R_{k_2\nu_{2}}^{2}}{(k_2^2+\nu_{2}^2)}
\frac{1}{(k_1^2+k_2^2+\nu_{1}^{2}+\nu_{2}^{2})}
\int\,d\tilde{\beta}(k)\frac{R_{k\nu}^{2}}{(k^2+\nu^2)}. \label{268}
\end{equation}
So, the regularized part of the contribution above reads:
\begin{equation}
I=\frac{1}{4\pi^{4}}\int_{0}^{\infty}\,d\nu\int_{0}^{\infty}\,d\vec{k}\int\,\frac{dk_{0}}{2\pi}
\nu \sinh(\pi\nu)K_{i\nu}(\mu z)K_{i\nu}(\mu z)
\frac{e^{\frac{-2\nu^{2}}{\Lambda^2}}}{k_{0}^{2}+\nu^{2}}.
\label{269}
\end{equation}
With some tedious manipulations \cite{prud} \cite{grad}
\cite{erdelyi}, we arrive at:
\begin{equation}
I =\biggl(\frac{\pi}{2}\biggr)^{1/2}\frac{\Lambda
m}{16}\,e^{\frac{\pi^{2}\Lambda^{2}}{8}}f(\Lambda, z), \label{270}
\end{equation}
where the function $f(\Lambda, z)$ is given by
\begin{equation}
f =
\int_{-\infty}^{+\infty}\,\frac{d\alpha}{\gamma}K_{1}(m\gamma)e^{-\frac{\Lambda^{2}\alpha^2}{8}}
\Biggl(
e^{\frac{i\pi\alpha\Lambda^{2}}{4}}erfc\biggl(-\frac{\Lambda}{2\sqrt{2}}(i\alpha+\pi)\biggr)
-
e^{-\frac{i\pi\alpha\Lambda^{2}}{4}}erfc\biggl(-\frac{\Lambda}{2\sqrt{2}}(i\alpha
-\pi)\biggr)\Biggr), \label{271}
\end{equation}
where $\gamma^{2} = 2z^{2}(1+\cosh\alpha)$ and $erfc(x)$ is the
complementary error function, which satisfies the identity:
\begin{equation}
erfc(-x) = 2 - erfc(x).
\end{equation}

It is not possible to present the solution of the integral in
Eq.(\ref{271}) in terms of known functions. In spite of this
inconvenience, it is easy to see that this function $f(\Lambda,
z)$ has strong convergence, proving that we indeed regularize the
contribution for the two-point function at the one-loop level.
Similar calculations can be carried out for the four-point
function. From the discussions it should be clear that is a simple
matter to derive the Schwinger functions of the theory at least in
the one-loop level.

There are two rather subtle points that we have to investigate.
The first one is related to the behavior of the random noise near
event horizon, and the second one is the validity of the program
beyond the Euclidean signature.

\section{Conclusions and perspectives}

\quad {\,\,\,}The stochastic quantization method was used to study
self-interacting fields in manifolds which can be analytically
continued to the Euclidean situation, i.e., static Riemannian
manifolds, namely, the Einstein manifold and the Rindler manifold.
First, we have solved a Langevin equation for the mode
coefficients of the field, then we exhibit the two-point function
at the one-loop level. It was shown that it diverges and we have
used a covariant stochastic regularization to regularize it. It
was shown that, indeed, the two-point function is regularized.

A natural question that arises, when we work in Rindler space, is
that what happens to the noise field correlation function, given
by Eq.(\ref{212}), near an horizon. From this equation, we see
that, whenever we have $g = \det g_{\mu\nu} = 0$, this correlation
function diverges, and, therefore, all n-point correlation
functions
$\langle\,\varphi(\tau_{1},x_{1})\varphi(\tau_{2},x_{2})...
\varphi(\tau_{n},x_{n})\,\rangle_{\eta}$ will have meaningless
values, in virtue of the solution of the Langevin equation. We may
implement a brick wall-like model \cite{thooft1} \cite{thooft2} in
order to account for these effects; in other words, we may impose
a boundary condition on solutions of the Langevin equation at a
point near the horizon. On the other hand, in the limit $g = \det
g_{\mu\nu}\rightarrow\infty$, all the correlation functions
vanish. We would like to point out that another and powerful way
to understand the behavior of the correlation functions in the
limit $g = \det g_{\mu\nu} = 0$ may be the stochastic quantization
of the two-dimensional Schwinger model in a curved background.
This can also be a way to study the most pertinent form of the
Langevin equation to be used in curved spacetimes. This subject
and also the link between our results and the program to implement
the perturbative renormalization in domains where the
translational invariance is broken \cite{boun1} \cite{boun2}
\cite{boun3} \cite{aparicio} is under investigation by the
authors.

We discussed also that the presence of the Markov parameter as an
extra dimension lead us to a regularization scheme, which preserves
all the symmetries of the theory under study. This is a quite
important point in a regularization procedure. It must preserve all
the symmetries of the unregularized Lagrangian. Although many
authors have stressed that we can not expect that a regularization
independent proof of the renormalization of theories in a curved
background exists, since the stochastic regularization is not an
action regularization, may be a way to construct such proof. We are
aware of the fact that the stochastic quantization program can be
implemented without problems (modulo event horizon problems, etc),
if it is possible to perform the Wick rotation, obtaining a real
Euclidean action.

The picture that emerges from the discussions of the paper is that
the implementation of the stochastic quantization in curved
background is related to the following fact.  For static manifold,
it is possible to perform a Wick rotation, i.e., analytically
extend the pseudo-Riemannian manifold to the Riemannian domain
without problem. Nevertheless, for non-static curved manifolds we
have to extend the formalism beyond the Euclidean signature, i.e.,
to formulate the stochastic quantization in pseudo-Riemannian
manifold, not in the Riemannian space (as in the original
Euclidean space) as was originally formulated. See for example the
Refs. \cite{con1} \cite{gozzi}. In the Ref. \cite{con1}, the
authors proposed a modification of the original Parisi-Wu scheme,
introducing a complex drift term in the Langevin equation, to
implement the stochastic quantization in Minkowski spacetime.
Gozzi \cite{gozzi} studied the spectrum of the non-self-adjoint
Fokker-Planck Hamiltian to justify this program. See also the
Refs. \cite{con2} \cite{con3}. Of course, this situation is a
special case of ordinary Euclidean formulation for systems with
complex actions.

The main difference between the original Euclidean and the
implementation of the stochastic quantization in Minkowski spacetime
is the fact that in the usual case a real Euclidean action the
approach to the equilibrium state is a stationary solution of the
Focker-Planck equation. In the Minkowski formulation, the
Hamiltonian is non-Hermitian and the eigenvalues of such Hamiltonian
are in general complex value. The real part of such eigenvalues are
important to the asymptotic behavior at large Markov time. The
approach to the equilibrium is achieved only if we can show its
positive semi-definiteness. The fundamental question is: what
happens if the Langevin equation describes diffusion around complex
action? Some authors claim that it is possible to obtain meaningful
results out of Langevin equation diffusion processes around complex
action. See for example the seminal paper of Parisi \cite{con4}.
Klauder et al. \cite{con6} investigated the complex Lagevin
equation, where some numerical simulations in one-dimensional
systems was presented. See also the Refs. \cite{con5} \cite{con7}.
Finally we would like to mention the approach developed by Okamoto
et al. \cite{okamoto}, where the role of the kernel in the complex
Langevin equation was studied.

We would like to remark that there are many examples where Euclidean
action is complex. The stochastic quantization in Minkowski
spacetime, as we discussed; system with chemical potential as for
example $QCD$ with non-vanishing chemical potential at finite
temperature; for $SU(N)$ theories with $N>2$, the fermion
determinant becomes complex and also the effective action. Complex
terms can also appears in the Langevin equation for fermions, but a
suitable kernel can circumvent this problem. Another quite
instructive case that deserves our attention is the stochastic
quantization of topological field theory. One of the peculiar
feature within this kind of theory is the appearance of a factor $i$
in front of the topological action in Euclidean space. In a
topological theory, the path integral measure weighing remains to be
$e^{iS}$, even after the Wick rotation. An attempt to use a
Markovian Langevin equation with a white noise to quantize the
theory, fails since the Langevin equation will not tend to any
equilibrium at large Markov parameter. In the literature there are
different proposed to solve the above mentioned problem. In a pure
topological Chern-Simons theory, Ferrari et al. \cite{con8}
introduced a non-trivial kernel in the Langevin equation. Other
approach was developed by the Menezes et al. \cite{menezes1}. These
authors showed that using a non-Markovian Langevin equation with a
colored random noise, the convergence problem can be solved. These
authors proved that it is possible to obtain convergence toward
equilibrium even with an imaginary Chern-Simons coefficient. We
conclude saying that several alternative methods have been proposes
to deal with interesting physical systems where the Euclidean actio
is complex. These methods do not suggest any general way of solve
the particular difficulties that arise in each situation.

The program of using the complex Langevin equation for an
investigation of quantum field theories in Minkowski spacetime is
not yet established and is still under discussion. Its clear that
until the problem that we discussed above is solved, one can not say
for sure that the implementation of the program of stochastic
stochastic quantization in non-static pseudo-Riemannian manifolds
can be successfully implemented.

\section{Acknowlegements}

We would like to thank  S. Joffily and G. Flores-Hidalgo for
enlightening discussions. This paper was supported by Conselho
Nacional de Desenvolvimento Cientifico e Tecnol{\'o}gico do Brazil
(CNPq).

\begin{appendix}
\makeatletter \@addtoreset{equation}{section} \makeatother
\renewcommand{\theequation}{\thesection.\arabic{equation}}

\section{The Kruskal extension of the Schwarzschild
spacetime.}

The spacetime of a spherically symmetric non-rotating and
eletrically neutral body is described by the Schwarzschild metric
\begin{equation}
ds^{2}=-\left(1-\frac{2M}{r}\right)dt^{2}+
\left(1-\frac{2M}{r}\right)^{-1}dr^{2}+r^{2}d\Omega^{2},
\label{a1}
\end{equation}
where $d\Omega^{2}=d\theta^{2}+\sin^{2}\theta\,d\phi$ is the
metric of the unit two-sphere. The coordinate system
$(t,r,\theta,\phi)$ provides a frame in which the metric
components are time-independent. Since, it is possible to find a
Killing vector field which is hypersurface orthogonal to the
family of spacelike hypersurfaces $t=cte$, the Schwarzschild
solution is also static. The line element given by  Eq.(\ref{a1})
describes the external gravitational field generated by any
spherical mass, whetever its radius \cite{bir}. This includes
several interesting situations, as a spherically symmetric star
which undergoes radial pulsations or ever a radial spherically
symmetric gravitational collapse. In the extreme case of a
complete gravitational collapse, we have to consider the
Schwarzschild line element as an empty spacetime solution for all
values of r. Since the line element has two singularities, one at
$r=0$ and another at $r=2M$, it represents only one of the patches
$0<r<2M$ or $2M<r<\infty$. It is not difficult to show that
although the Schwarzschild line element is singular at $r=2M$, all
the invariants constructed with the Riemann tensor and its
contractions are well-behaved at $r=2M$, and this point is a
singularity due to an inappropriate choice of coordinates.

On the other hand, choosing the Schwarzschild line element to
describes the patch $0<r<2M$, we have that the curvature scalar
diverges at $r=0$, since we have
$R_{\mu\nu\rho\sigma}R^{\mu\nu\rho\sigma}=\frac{48M^{2}}{r^{6}}$.
Therefore this point is a real spacetime singularity. Thus,
although it is not possible to extend the spacetime across the
singularity, it is posssible to analytically continue the manifold
from $r>2M$ to the region  $r<2M$, $(r\neq 0)$. The maximal
extension of the manifold described by the Schwarzschild line
element with $2M<r<\infty$ was obtained by Kruskal.

From the condition for null geodesics it is possible to show that
the radial null geodesics in the Schwarzschild spacetime are given
by
\begin{equation}
\left(\frac{dt}{dr}\right)^{2}= \left(\frac{r}{r-2M}\right)^{2},
\label{a2}
\end{equation}
and allow us to define $r^{*}$ as
\begin{equation}
r^{*}=\int\frac{dr}{\left(1-\frac{2M}{r}\right)}=
r+2M\ln\left(\frac{r}{2M}-1\right), \label{a3}
\end{equation}
and it is clear that the radial geodesics must satisfy
$t=r^{*}+cte$ or $t=-r^{*}+cte$.

The next step is to define the null coordinates $u$ and $v$, where
$u=t-r^{*}$ and $v=t+r^{*}$. Obviously, it is possible to write
the Schwarzschild line element in terms of the null coordinates
$u$ and $v$, and we have
\begin{equation}
ds^{2}=-\frac{2Me^{-\frac{r}{2M}}}{r}\,e^{\frac{(v-u)}{4M}}dudv+
r^{2}d\Omega^{2}, \label{a4}
\end{equation}
where the metric of the unit two-sphere is inalterated. To go
further, let us define new coordinates $U$ and $V$, where
\begin{equation}
U=\exp(\frac{u}{4M}), \label{a5}
\end{equation}
and
\begin{equation}
V=\exp(\frac{v}{4M}), \label{a6}
\end{equation}
Using the coordinates $U$ and $V$, the metric can be written as
\begin{equation}
ds^{2}=-\frac{32M^{3}e^{-\frac{r}{2M}}}{r}dUdV+r^{2}d\Omega^{2}.
\label{a7}
\end{equation}
The value $r=2M$ is no more a singularity, since corresponds to
$U=0$ or $V=0$. Finally, to obtain the Schwarzschild  metric in
the Kruskal form we have only to define the coordinates $T$ and
$X$, by choosing $T=\frac{1}{2}(U+V)$ and $X=\frac{1}{2}(V-U)$.
The final form of the  Schwarzschild  metric in terms of the
Kruskal coordinates $(T,X,\theta,\phi)$ is
\begin{equation}
ds^{2}=\frac{32M^{3}e^{-\frac{r}{2M}}}{r}(-dT^{2}+dX^{2})
+r^{2}d\Omega^{2}. \label{a8}
\end{equation}
Note that the coordinate transformation between the original
coordinates $(t,r)$ and the Kruskal coordinates $(T,X)$ is given
by
\begin{equation}
X^{2}-T^{2}=\left(\frac{r}{2M}-1\right)\exp\left(\frac{r}{2M}\right),
\label{a9}
\end{equation}
and
\begin{equation}
t=4M\tanh^{-1}\left(\frac{T}{X}\right). \label{a10}
\end{equation}

Therefore, we show how it is possible to analytically continue the
manifold from $r>2M$ to the region $r<2M$, and the event horizon,
a sphere of radius $r=2M$ is only a coordinate singularity, which
can be removed by a suitable coordinate transformation. Althought
the apparent singularity at the horizon has disappeared, there are
true singularities in the Kruskal extension of the Schwarzschild
spacetime. The physical singularity at $r=0$ corresponds to the
values $X=\sqrt{T^{2}-1}$ and $X=-\sqrt{T^{2}-1}$. The original
Schwarzschild solution for $r>2M$ corresponds to the region where
observers can obtain information. Since there are two event
horizon, the future event horizon and the past event horizon, the
Kruskal spacetime defines two outer asymptotically flat regions
and also two regions inside the event horizon, bounded by the past
and future singularities. The black-hole is the maximal analytic
extension of the vacuum Schwarzschild solution. It is well known
that close to the horizon, the Schwarzschild coordinates $t$ and
$r$ behaves as Rindler's spacetime coordinates.

\end{appendix}

\end{document}